\documentclass{aastex62}

\newcommand{\au}{{\rm\,AU}}

\received{\today}
\revised{}
\accepted{}
\submitjournal{ApJ}

\shorttitle{The central 1000 AU of a pre-stellar core}
\shortauthors{Caselli et al.}
\begin{document}

\title{The central 1000 AU of a pre-stellar core revealed with ALMA. I. 1.3\,mm continuum observations}

\correspondingauthor{Paola Caselli}
\email{caselli@mpe.mpg.de}

\author{Paola Caselli} 
\affil{Max-Planck-Institut f\"ur Extraterrestrische Physik, Gie{\ss}enbachstrasse 1, 85741 Garching bei M\"unchen, Germany} 

\author{Jaime E. Pineda}
\affil{Max-Planck-Institut f\"ur Extraterrestrische Physik, Gie{\ss}enbachstrasse 1, 85741 Garching bei M\"unchen, Germany} 
%\nocollaboration

\author{Bo Zhao}
\affil{Max-Planck-Institut f\"ur Extraterrestrische Physik, Gie{\ss}enbachstrasse 1, 85741 Garching bei M\"unchen, Germany} 

\author{Malcolm C. Walmsley}
\altaffiliation{Deceased}
\affiliation{INAF-Osservatorio Astrofisico di Arcetri, Largo E. Fermi 5, 50125 Firenze, Italy}
\affiliation{Dublin Institute of Advanced Studies, Fitzwilliam Place 31, Dublin 2, Ireland}
%\nocollaboration

\author{Eric Keto}
\affiliation{Harvard-Smithsonian Center for Astrophysics, 160 Garden Street,  Cambridge, MA 02420, USA}
%\nocollaboration

\author{Mario Tafalla}
\affiliation{Observatorio Astron\'omico Nacional (OAG-IGN), Alfonso XII 3, 28014, Madrid, Spain}

\author{Ana Chac\'on-Tanarro}
\affil{Max-Planck-Institut f\"ur Extraterrestrische Physik, Gie{\ss}enbachstrasse 1, 85741 Garching bei M\"unchen, Germany} 
\affiliation{Observatorio Astron\'omico Nacional (OAG-IGN), Alfonso XII 3, 28014, Madrid, Spain}

\author{Tyler L. Bourke}
\affiliation{SKA Organization, Jodrell Bank Observatory, Lower Withington, Macclesfield, Cheshire SK11 9DL, UK }

\author{Rachel Friesen}
\affiliation{National Radio Astronomy Observatory, 520 Edgemont Rd, Charlottesville, VA 22903, USA}

\author{Daniele Galli}
\affiliation{INAF-Osservatorio Astrofisico di Arcetri, Largo E. Fermi 5, 50125 Firenze, Italy}

\author{Marco Padovani}
\affiliation{INAF-Osservatorio Astrofisico di Arcetri, Largo E. Fermi 5, 50125 Firenze, Italy}



\begin{abstract}

Stars like our Sun form in self-gravitating dense and cold structures within interstellar clouds, called pre-stellar cores. Although much is known about the physical structure of dense clouds just before and soon after the switch-on of a protostar, the central few thousand astronomical units (au) of pre-stellar cores are unexplored. It is within these central regions that stellar systems assemble and fragmentation may take place, with the consequent formation of binaries and multiple systems.  We present ALMA Band 6 observations (ACA and 12m array) of the dust continuum emission of the 8\,M$_{\odot}$ pre-stellar core L1544, with angular resolution of 2$^{\prime\prime}\times1.6^{\prime\prime}$  (linear resolution 270\,au\,$\times$\,216\,au).  Within the primary beam, a compact region of 0.1\,M$_{\odot}$, which we call a "kernel", has been unveiled. The kernel is elongated, with a central flat zone with radius $R_{\rm ker}\simeq$10$^{\prime\prime}$ ($\simeq$1400\,au). The average number density within $R_{\rm ker}$ is $\simeq$1$\times$10$^{6}$\,cm$^{-3}$, with possible local density enhancements.  The region within $R_{\rm ker}$ appears to have fragmented, but detailed analysis shows that similar substructure can be reproduced by synthetic interferometric observations of a smooth centrally concentrated dense core with a similar central flat zone.  The presence of a smooth kernel within a dense core is in agreement with non-ideal magneto-hydro-dynamical simulations of a contracting cloud core with a peak number density of 1$\times$10$^7$\,cm$^{-3}$.  Dense cores with lower central densities are completely filtered out when simulated 12m-array observations are carried out. These observations demonstrate that the kernel of dynamically evolved dense cores can be investigated at high angular resolution with ALMA. 
%When combined with spectroscopic data and higher sensitivity continuum observations, they will furnish important information on the structure and kinematics of cloud cores on the verge of star formation. 
\end{abstract}

\keywords{stars: formation --- ISM: clouds --- techniques: image processing, interferometric --- (magnetohydrodynamics:) MHD --- methods: numerical}


\section{Introduction} \label{sec:intro}

Dense cloud cores represent the initial conditions in the process of individual star formation \citep{SAL87}. Prior to the formation of young stellar objects in their central regions, these objects are called starless cores. They accrete material from the surrounding less dense cloud and a fraction of them eventually become gravitationally unstable, entering the pre-stellar core phase \citep{ADW14}.   Pre-stellar cores contract under the pull of gravity, only counteracted by magnetic and thermal pressure, with small contribution from turbulence \citep{GBW98}.  They can form single \citep{EDL15} or multiple \citep{POP15} stellar systems depending on the interplay between magnetic fields and the partially ionised material present within these clouds \citep{CHH11,BL13}. Detailed dynamical models of pre-stellar core evolution have been built, showing that the contraction of rotating and magnetised cold cores is a complex process which does not necessarily lead to the formation of a protoplanetary disk \citep{LBP14} or multiple stellar systems \citep{RCB14}. To shed light on pre-stellar core evolution and the formation of multiple stellar systems, we ultimately need observational constraints for model predictions. In particular, to study the structure and gas motions within the central 1000\,au of a pre-stellar core requires high sensitivity observations of dust continuum emission and molecular lines at size scales similar to those of protoplanetary disks ($\le$200\,au).

Several attempts have been made in the past to detect the central regions and possible substructure of dense cloud cores before star formation (i.e. starless cores, or, if self-gravitating, pre-stellar cores), but no positive results have been collected so far. No millimetre dust continuum emission was detected toward the pre-stellar core L1544 in the Taurus Molecular Cloud Complex, using the IRAM Plateau the Bure Interferometer \citep{CCW07}. The conclusion was that, if an early-stage protostar is present in the center of L1544, its mass cannot exceed 0.2 Jupiter masses (M$_{\rm J}$). No dust continuum detection was found within 11 starless cores in the Perseus Molecular Cloud, using CARMA, the Combined Array for Research in Millimeter-wave Astronomy \citep{SEJ10}. The implication was that Perseus starless cores are characterised by shallow density profiles in the inner few thousand au. Other interferometric observations (with CARMA and SMA, the Submillimeter Array) have also been carried out to investigate the possible presence of substructure toward five bright starless cores in Perseus and Ophiuchus \citep{SSD12}. Once again, the non-detections implied that the observed cores have flat density profiles out to at least 1200\,au, but no images of these central regions could be recovered. 

\citet{BMC12} used the SMA to study the dense starless core N6 in the nearest cluster forming region, the $\rho$Oph\,A cloud in the Ophiuchus Molecular Cloud Complex, and found a poorly resolved compact source with size $\sim$1000\,au and mass 0.005-0.01\,M$_{\odot}$, not seen with single dish observations. A compact structure with mass $\leq$20\,M$_{\rm J}$ was detected toward the starless core SM1N in the same cluster forming region $\rho$Oph\,A with the Atacama Large Millimetre and sub-millimetre Array (ALMA) in Cycle 1 \citep{FDB14}, but the sensitivity was not high enough to study its structure. More recent observations \citep{FPB18} have shown that both SM1N and N6 may be at a later evolutionary stage, with presence of compact dust continuum emission and possible evidence of CO outflow. Another starless core, L1689N in Ophiuchus, has been imaged with the Atacama Compact Array (ACA, also known as Morita Array), by \citet{LWG16}, who found a compact dust continuum source with size similar to N6 but a mass of 0.2-0.4\,M$_{\odot}$; however, the ACA angular resolution was not high enough to study the inner structure. Moreover, L1689N is located on the East side of the Class\,0 binary source IRAS\,16293-2422 and it is impacted by the outflow driven by one of the protostars, so its structure and kinematics have been probably affected by the nearby star formation activity. Still focusing on the actively star-forming Ophiuchus complex, \citet{KDD17} carried out ALMA Cycle 2 observations of 60 starless and protostellar cores and only detected one starless core, without spatially resolving it.  

Cluster forming regions, such as Ophiuchus and/or active sites of star formation, tend to have complex structure, with clear presence of multiple dense cores within a higher density medium \citep[e.g.][]{FDB14, WZT14, HCF16, OSS18}; however, the study of these regions does not allow us to quantify how much of the observed structure is due to the original dynamical evolution of the pristine cloud, i.e. before star formation, and how much of such structure is instead a result of protostellar feedback. More quantitative work on the {\em initial conditions} of the process of star formation, including the detailed inner structure of isolated dense cores and their possible fragmentation, is better done in regions not affected by feedback, such as isolated dense cores in molecular clouds (e.g. the pre-stellar core L1544 in Taurus, studied here) or Bok Globules, although this is a difficult task.  For example, recent work with (Cycle 1) ALMA toward starless cores in the Chamaeleon Molecular Cloud \citep{DOP16} failed to detect any of the 56 starless cores in their sample. Besides the high sensitivity needed to observe these cold regions, which now is available with ALMA, another problem associated with the detection of nearby starless cores is that only those with high volume densities within compact central regions, or "kernels"  \citep[extending the definition used by][for cluster-forming molecular cloud cores, to low-mass pre-stellar cores]{M98} are expected to be revealed by interferometers; these centrally concentrated cores on the verge of star formation (the so-called pre-stellar cores) are dynamically evolved and have short life times \citep{ADW14}. To summarise, pre-stellar cores are rare and difficult to find. 

Here we present the ALMA (12\,m array and ACA) 1.3\,mm dust continuum emission map of the 8 M$_{\odot}$ pre-stellar core L1544 in Taurus at a distance of 135\,pc \citep[][; consistent with the recent Taurus distance measurements of \cite{GLO18}]{SGF14}, which for the first time resolves its central 1000\,au. In past work, L1544 has been modelled as a Bonnor-Ebert (BE) sphere \citep{B56,E57} in quasi-static contraction \citep{KC10,KRC14}. The BE sphere has a central region, called the flat zone, with radius $R_{\rm flat}$, where the volume density is constant, and which can be theoretically determined by multiplying the sound speed with the free-fall time at the central density \citep{KC10}.

The observations are described in Section\,2, observational results are in Section\,3, the comparison with smooth cloud models is in Section\,4, hydro- and magneto-hydro simulations of contracting pre-stellar cores are presented in Section\,5, discussion and conclusions are in Section\,6. 

\section{Observations} \label{sec:observations}

We conducted ALMA observations of the L1544 pre-stellar core during Cycle 2 (ESO Project ID 2013.1.01195.S, PI Caselli). Both the 12-m (Main array) and the 7-m (ACA) arrays were employed. The 12-m array observations were carried out on 2014 Dec 27 and 29, while the 7-m array observations were carried out on 2014 June 15, July 20 and 29, and 
August 6 and 11.  The single pointing observations used a correlator configuration where 1 of the 4 correlator spectral windows was used for continuum centred at 228.973\,GHz and a 2\,GHz bandwidth, while the other 3 spectral windows were used for N$_2$D$^+$, DCO$^+$, and D$^{13}$CO$^+$. These data will be presented in a follow-up paper dedicated to the kinematics (Paper II). 

The 12-m array observations used the quasar J0423-020 and J0510+1800 as bandpass and gain calibrators, respectively, while flux calibration was done using Uranus. The 7-m array observations used the quasar J0510+1800 as bandpass and gain calibrator, while flux calibration was done using J0510+1800, Ganymede, and Uranus observations. The 12-m array data were calibrated using the Common Astronomy Software Applications package \citep[CASA;][]{MWS07} version 4.2.2, while the 7-m array data were calibrated using CASA version 4.2.1. Before combining the data from both arrays we calculated the relative weights using the {\em statwt()} command. We perform a joint de-convolution of both datasets (ACA and 12-m array) in CASA (version 4.4.0) using multifrequency synthesis ({\em mfs}), a robust weighting of 0.5, and a taper of 1.5$^{\prime\prime}$, which results in a beam with major and minor axes of 2.05$^{\prime\prime}\times1.61^{\prime\prime}$ (277\,au $\times$ 217\,au at the distance of Taurus) with a position angle of 59.5$^{\circ}$. Since the continuum emission is extended, we use the multi-scale clean technique, with the multi-scale parameter of [0, 2, 6, 18]\,arcsec. The noise level was estimated using the root mean square over an emission free section of the image and found to be 36\,$\mu$Jy\,beam$^{-1}$.

\section{Results} \label{sec:results}

\begin{figure}[ht!]
\includegraphics[width=\textwidth]{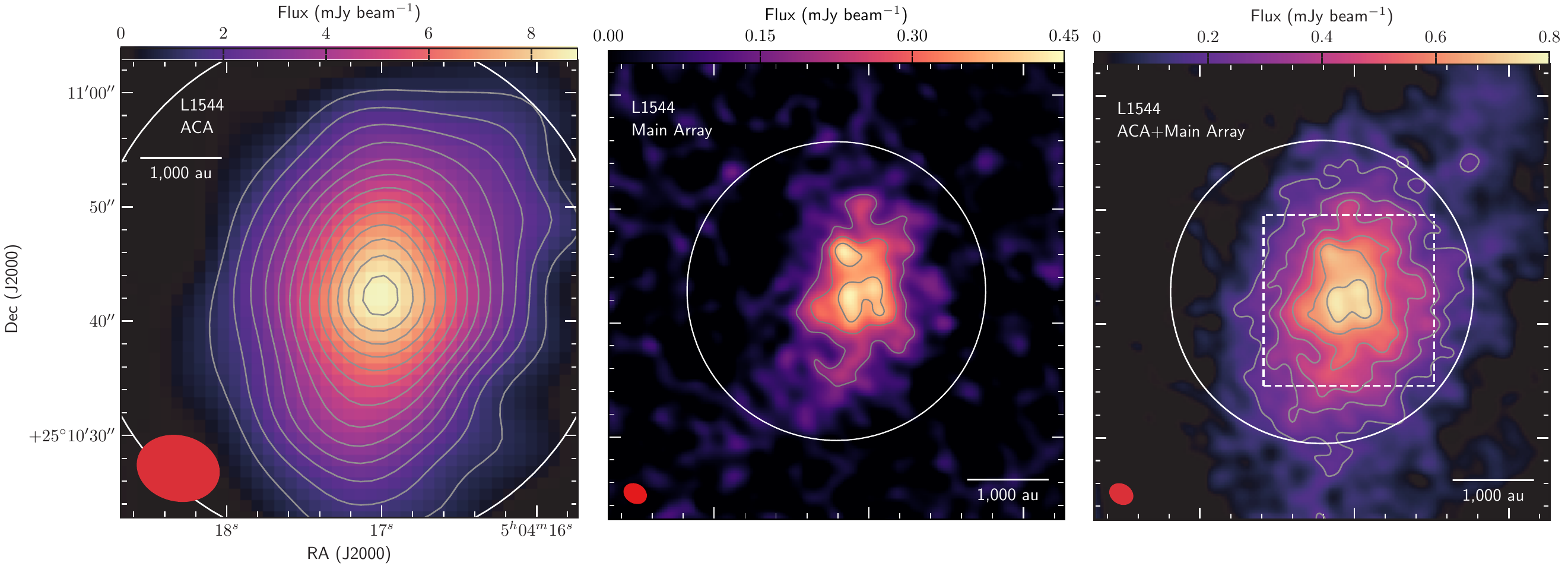}
\caption{1.3mm dust continuum emission toward L1544 as seen by: (Left) the ACA, with contours starting at 5\,$\sigma$ (with $\sigma$ = 0.21\,mJy\,beam$^{-1}$) in steps of 3\,$\sigma$; (Center) the 12m-array only, with contours starting at 5\,$\sigma$ (with $\sigma$ = 34\,$\mu$Jy\,beam$^{-1}$) in steps of 3\,$\sigma$; (Right) combined ACA + 12m array, with contours starting at 5\,$\sigma$ (with $\sigma$ = 36\,$\mu$Jy beam$^{-1}$) in steps of 3\,$\sigma$. The white dashed square is the area shown in Figure\,\ref{fig:map_continuum}. The white circles in the three panels represent the corresponding 50\% primary beam response. The red ellipses in the bottom left of the three panels are the synthesized beams.}\label{fig:maps_array}
\end{figure}

\begin{figure}[ht!]
\plotone{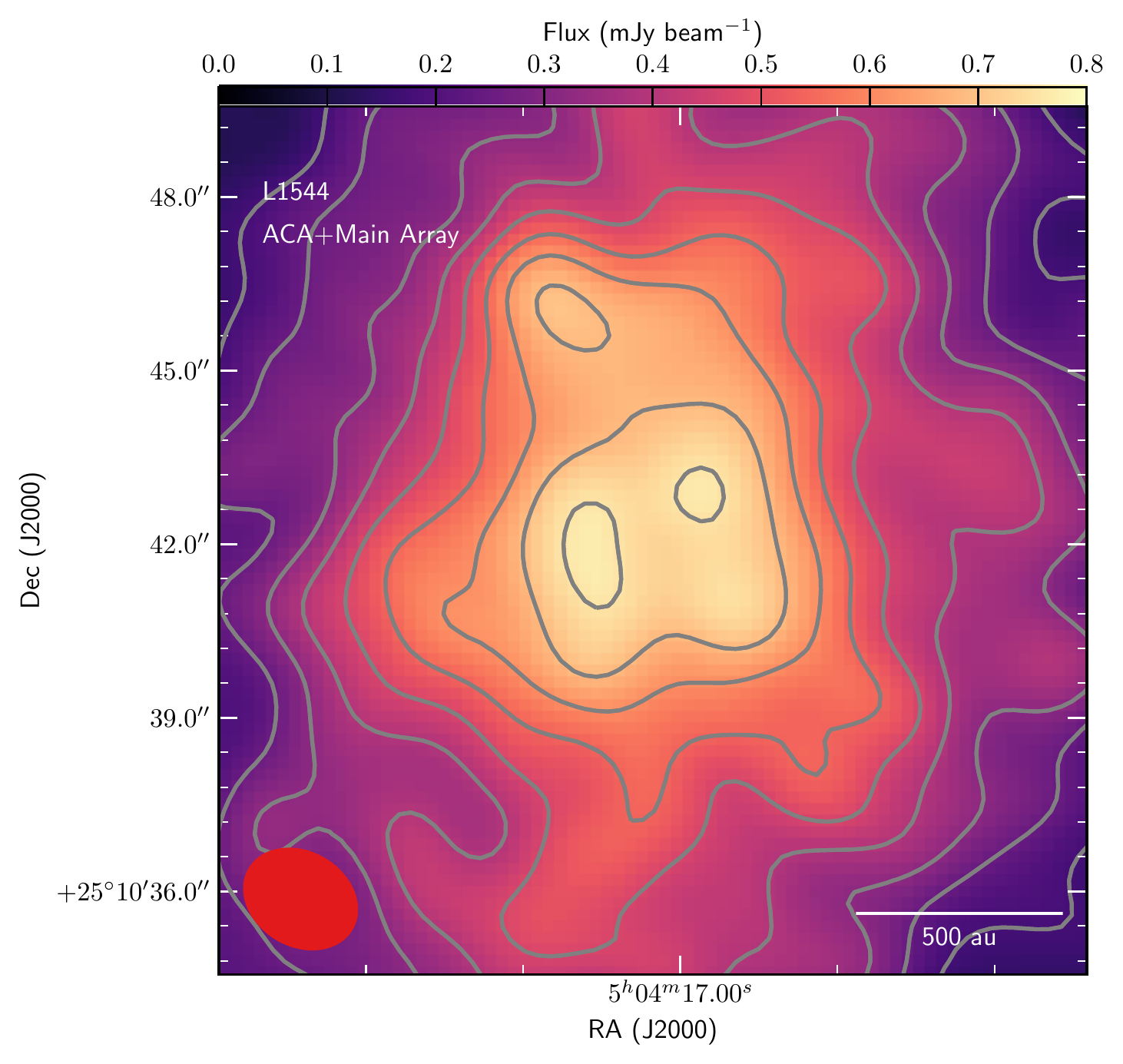}
\caption{Magnification of the central region of L1544 enclosed by the white dashed square in Fig.\ref{fig:maps_array}, right panel. The contour levels start from 5\,$\sigma$, in steps of 2\,$\sigma$, where $\sigma$ = 36\,$\mu$Jy\,beam$^{-1}$ (note that in Fig.\ref{fig:maps_array} the contour step was 3\,$\sigma$) . The synthesized beam is shown at the bottom left corner (red ellipse), and the scale bar is shown at the bottom right corner. Three separated fragments are seen in the central 1000\,au at a 2\,$\sigma$ level. \label{fig:map_continuum}}
\end{figure}

Figure\,\ref{fig:maps_array} shows the ACA-only dust continuum emission map of L1544 (left panel), the 12-m (Main) array map (central panel), and the ACA+Main array map (right panel), all without primary-beam correction to maintain a roughly constant noise across the image. Figure\,\ref{fig:map_continuum} presents a magnification of the combined ACA+Main Array map of the 1.3 mm dust continuum emission. 

Figures\,\ref{fig:maps_array} and \ref{fig:map_continuum} show that the high density kernel can now be viewed with unprecedented detail.  The kernel radius ($R_{\rm ker}$), derived from the area within the 5\,$\sigma$ contour of the combined map is 10.3\arcsec \ (1446\,au). Assuming a dust temperature $T_d$ = 6.5 K \citep{CCW07,KC10} and a dust opacity at 1.3 mm $\kappa_{\rm 1.3mm}$ = 0.009\,cm$^2$\,g$^{-1}$ \citep[corresponding to dust grains with thick icy mantles and 10$^6$\,yr of coagulation;][]{OH94}, consistent with recent findings \citep{CPC19}, the kernel mass can be found using: 

\begin{equation}
M_{\rm ker} = \frac{D^2 S_{\rm 1.3mm}}{\kappa_{\rm 1.3mm} B_{\rm 1.3mm}(T_d)},
\end{equation}
where D is the distance of the source (135 pc), $S_{\rm 1.3mm}$ is the primary-beam-corrected flux integrated within the 5\,$\sigma$ contour and within the 50\% primary beam response in the right panel of Figure\,\ref{fig:maps_array} (42$\pm$2\,mJy, taking into account a 5\% calibration error), and $B_{\rm 1.3mm}(T_d)$ is the Planck function at temperature $T_d$. Substituting the numerical values, we find $M_{\rm ker}$ = 0.100$\pm$0.005 M$_{\odot}$\footnote{The 1\,$\sigma$ error of the mass only contains the 1\,$\sigma$ error of the observed flux; however, uncertainties on the source distance and on the dust opacity effectively increase this error to about 50\,$\%$ of the measured value.}. The average H$_2$ number density is then $n({\rm H_2})$ = 3\,$M_{\rm ker}$/($4 \pi \mu_{\rm H_2} m_{\rm H} R_{\rm ker}^3$) = (1.00$\pm$0.05)$\times$10$^{6}$\,cm$^{-3}$, where $\mu_{\rm H_2}$ is the molecular weight per hydrogen molecule \citep[=2.8;][]{KBB08} and $m_{\rm H}$ is the H-atom mass; this $n({\rm H_2})$ value is in agreement with previous work by \citet{CCW07}. It is interesting to note that the kernel is elongated close to North-South, unlike the surrounding dense core, whose major axis is tilted by about 45$^{\circ}$, as deduced by larger scale observations; we will come back to this misalignment in Section\,\ref{sec:modeling}.

Fragments are apparent at a 3-$\sigma$ level in the Main array image (central panel in Fig.\,\ref{fig:maps_array}) and at a 2-$\sigma$ level in the combined (ACA+Main Array) map (Fig.\,\ref{fig:map_continuum}). The fragments are not spatially resolved, i.e. their size is less than the ALMA synthesized beam of 2.05$^{\prime\prime}$ $\times$ 1.60$^{\prime\prime}$ (277\,au $\times$ 216\,au). Using the above temperature and dust opacity, the mass enclosed within the fragments (the 11\,$\sigma$ contour in the central panel of Fig.\,\ref{fig:maps_array}) is 0.003 M$_{\odot}$ (3 M$_{\rm J}$), i.e. each fragment is about 1\,M$_{\rm J}$. The fragments appear to be centrally concentrated structures, with peak number densities of around 5$\times$10$^6$\,cm$^{-3}$, assuming spherical symmetry and a radius of 92\,au\footnote{The equivalent radius within the 11\,$\sigma$ contour of Fig\,\ref{fig:maps_array}, central panel, is 276\,au, so, assuming the presence of 3 fragments, we have simply divided this number by 3.}. Therefore, the fragments could be local density enhancements, where material has currently assembled, but they cannot be close to gravitational collapse, as their size is significantly smaller that the size scale below which stable oscillations rather than gravitational collapse can occur. In fact, considering an isothermal cloud where gravity is only counteracted by thermal pressure, one can find the size scale below which stable oscillations rather than gravitational collapse will occur. This is the Jeans length $L_J$, defined by: 

\begin{equation}
L_J  = \sqrt{\frac{\pi c_s^2}{G \rho_0}}, 
\end{equation}
where $G$ is the gravitational constant, $\rho_0$ is the mass density and $c_s$ is the sound speed: 

\begin{equation}
 c_s = \sqrt{\frac{k_{\rm B}T}{\mu_{\rm p} {\rm m_H}}}, 
 \end{equation}
with $k_{\rm B}$ the Boltzmann constant, $T$ the gas temperature, $\mu_{\rm p}$ the mean molecular weight per free particle \citep[2.37;][]{KBB08}. Assuming a temperature of 6.5\,K and the mass density of the kernel of 4.5$\times$10$^{-18}$\,g\,cm$^{-3}$ (derived from the measured mass, 0.1\,M$_{\odot}$,  radius, 1446\,au, assuming spherical symmetry and uniform density), the Jeans length of the kernel is about 3200\,au, about 30 times larger than the size upper limit of the fragments.  However, the fragments may also be imaging artefacts due to extended emission viewed with an interferometer and in the next session we will explore this point. 

%This value is 4 times lower than the maximum central density of 2$\times$10$^7$\,cm$^{-3}$, predicted by spherically symmetric models constrained by single dish observations \citep{KC10}.

\section{Comparison with smooth cloud models} \label{sec:smooth-model}

\subsection{The centrally concentrated Bonnor-Ebert sphere}

\begin{figure}[ht!]
\plotone{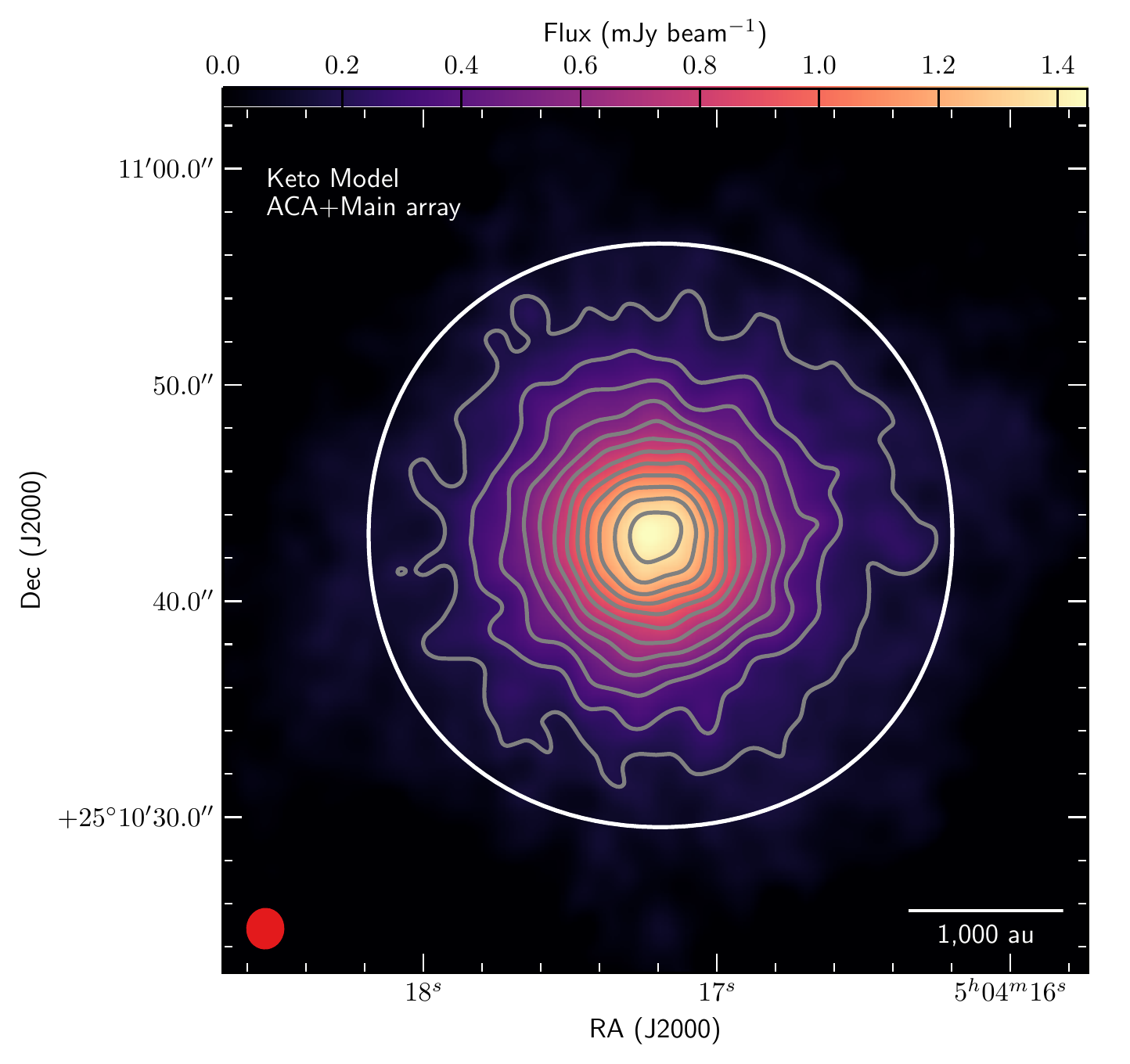}
\caption{Simulated ACA + 12m array observations of the Bonnor-Ebert sphere considered as the best-fit model of L1544 by \citet{KRC14}. Levels start at 5\,$\sigma$, in steps of 3\,$\sigma$ ($\sigma$ = 36\,$\mu$Jy\,beam$^{-1}$ is the noise level from observed data). The white circle is the 50\% primary beam response and the red ellipse in the bottom left corner is the synthesized beam. Comparison with the right panel of Figure\,\ref{fig:maps_array} shows that, within the central 1000\,au, this model is significantly more centrally concentrated than the observed pre-stellar core. \label{fig:Bonnor-Ebert}}
\end{figure}

We simulated similar ALMA (ACA and Main Array) observations to determine if the current best smooth model of the core could reproduce the observations. The model is generated using the physical parameters found by \citet{KC10} and refined by \citet{KRC14}, such that they reproduce the previous set of single dish observations of the dust continuum and molecular line emission. The structure deduced by \citet{KC10} and \citet{KRC14} is a Bonnor-Ebert sphere \citep{B56,E57}, with a temperature gradient which reproduces the one observed by \citet{CCW07}. The high-resolution continuum image is then processed using the task {\em simobserve} in CASA 4.6.0, which generates simulated data with a similar integration time and antenna configuration for both ACA and the Main array observations. The two datasets are then imaged using the same parameters as for the real observations, {\em briggs weighting} of 0.5, {\em taper} of 1.5$^{\prime\prime}$, and  multi-scale clean.  The resulting image (Figure\,\ref{fig:Bonnor-Ebert}) shows a core that is clearly more centrally concentrated than deduced by the observations, with a peak flux about two times larger than that observed. The central zone of a Bonnor-Ebert sphere, where the volume density is constant, is called the flat zone, and its radius is the flat radius, $R_{\rm flat}$; the flat zone is expected to shrink during the dynamical evolution of the core toward the formation of a protostar \citep[see e.g.][]{KC10}. L1544 has a larger flat radius than predicted by \citet{KC10}, thus suggesting that it is at an earlier evolutionary phase, or maybe that the L1544 dynamical evolution differs from that of a contracting Bonnor-Ebert sphere \citep[see, e.g., the three-dimensional L1544 structure derived by][using single dish dust continuum emission data]{DES05}, with possible formation of multiple fragments instead of a single peak.

\subsection{The smooth elongated structure derived from ALMA observations}   

The comparison between the observations (Fig.\,\ref{fig:maps_array}, right panel) and the Bonnor-Ebert model (Fig.\,\ref{fig:Bonnor-Ebert}) indicates that L1544 is not as centrally concentrated as predicted.  Here we consider a parametric model based on the smooth structure observed by ALMA, taking into account the elongated nature of the core to test further the existence of fragments within the core centre. For this, we fix the peak flux of the model to match the peak flux of the combined (ACA+Main Array) image; this clearly will underestimate the peak flux of the simulated combined image, as the more extended structure already filtered out by ALMA is not taken into account in the model.  However, this is a compromise to allow a better fit to the central 3000\,au, where most of the flux resides, compared to the outer regions which are better traced by single dish observations (see Section\,\ref{sec:chacon}). 

The following function is used to model the structure observed in Figure\,\ref{fig:maps_array}, right panel: 

\begin{equation}
F(r) = \frac{F_0}{1+(r/R_{\rm flat})^{\alpha}} ,
\end{equation}

\noindent
where $F(r)$ is the flux density at radius $r$, defined as the elliptical distance to the core center:  

\begin{equation}
r = \sqrt{x_{maj}^2 + (y_{min}\times AR )^2} ,
\end{equation}

\noindent
where $x_{maj}$ and $y_{min}$ are the coordinates of $r$ along the core major and minor axes, respectively, in a system of reference where the major axis is along the $x$-axis and the minor axis is along the $y$-axis, and $AR$ is the major to minor axis ratio of the whole modelled structure. The best-fit model is calculated by minimising the $\chi^2$ of the flux difference between the model (after applying the primary beam correction) and the combined ACA+Main Array data over the 50\% primary beam response region.  This includes core centre, orientation, axes ratio, and the radial profile parameters.   The best-fit model has the following parameters: $F_0$=0.772\,mJy\,beam$^{-1}$,  $R_{\rm flat}$=10.84$^{\prime\prime}$ (equivalent to 1463\,au), $\alpha$=2.32, $AR$=1.83, position angle PA=166$^{\circ}$, measured due East from North. The peak value is fixed and the minimisation is carried out over the region with primary beam response higher than 50\%.

\begin{figure}[ht!]
\includegraphics[width=\textwidth]{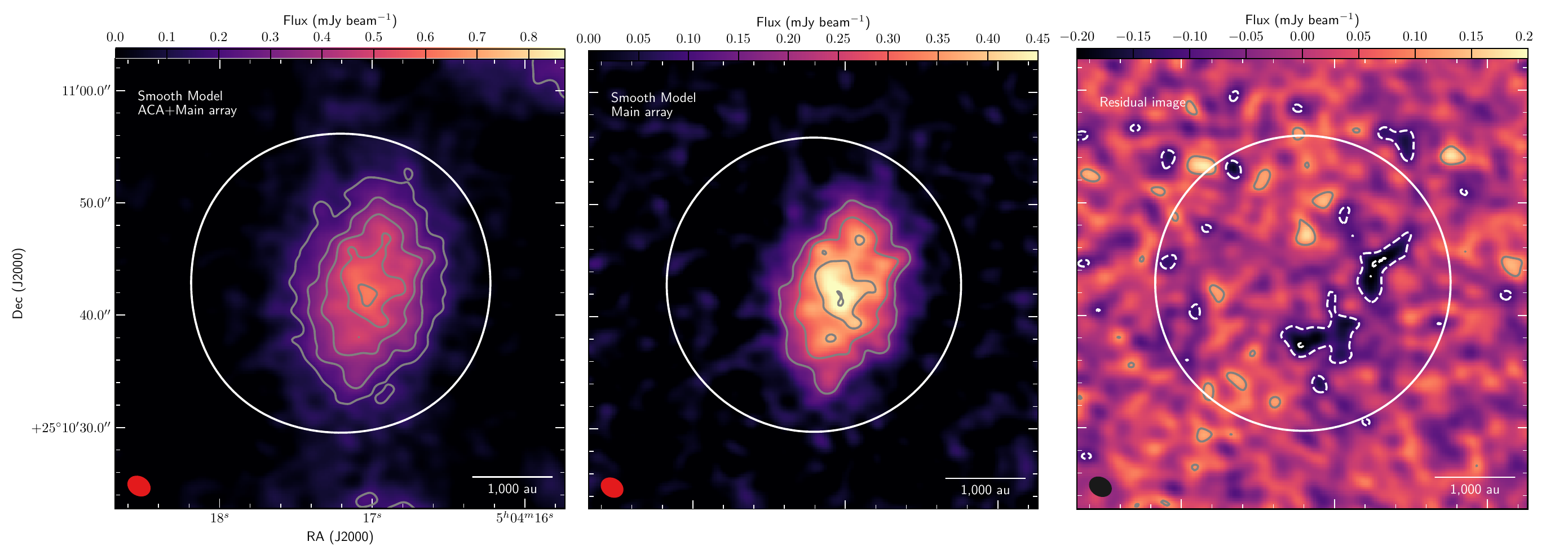} 
\caption{Results of simulating a model cloud with a density distribution obtained from the observations (right panel of Fig.\,\ref{fig:maps_array}). The color stretch and contour levels of the left and center panel are exactly the same (in Jy\,beam$^{-1}$) as those used for the data figures (starting from 5\,$\sigma$ and in steps of 3\,$\sigma$, with $\sigma$ = 36\,$\mu$Jy\,beam$^{-1}$). The left panel shows the ACA+Main Array observation simulations (to be compared with the right panel of Fig.\,\ref{fig:maps_array}); the central panel shows the Main array observation simulations (to be compared with the central panel of Fig.\,\ref{fig:maps_array}); the right panel is the residual image after subtracting the simulated image in the central panel to the equivalent data image in Fig.\,\ref{fig:maps_array}. The contours in this last panel start from 3\,$\sigma$ and they are in steps of 1\,$\sigma$. The dashed contours represent negative 3-$\sigma$ contours.}
\label{fig:model2}
\end{figure}

The model is then passed through the ALMA simulator observation and processed in the same way as for the Bonnor-Ebert model. Figure\,\ref{fig:model2} shows the results of this analysis.  The left panel shows the simulated ACA+12m array map, the central panel shows the simulated 12m-array-only observations and the right panel shows the residuals obtained from the subtraction of the data in the central panel of Fig.\,\ref{fig:maps_array} and the model in the central panel of Fig.\,\ref{fig:model2}.  Although the simulated image does not reproduce the observed substructure, the residual image does not show fragments at a significant level, thus implying that the observed substructure is not significant. The next subsection focusses on the longer baselines to highlight the compact emission and further test the smooth-elongated structure model. 

\subsubsection{The smooth elongated structure observed with baselines longer than 13\,k$\lambda$ \label{App:substructure}}
We now show the comparison between data and simulated observations of the smooth elongated structure model when only baselines longer than 13\,k$\lambda$ are used, to give more weight to the substructure.   With the aim of confirming or discarding the presence of substructure within the central 1000\,au of the pre-stellar core L1544, we generate synthetic ALMA observations using an array configuration similar to the one used during the real observations (Configuration 2 of Cycle 1 included in CASA) and the same on-source time using the {\em simobserve} task in CASA. We run 100 models with different starting hour-angles to generate different $uv$-coverage.

\begin{figure}[ht!]
\plotone{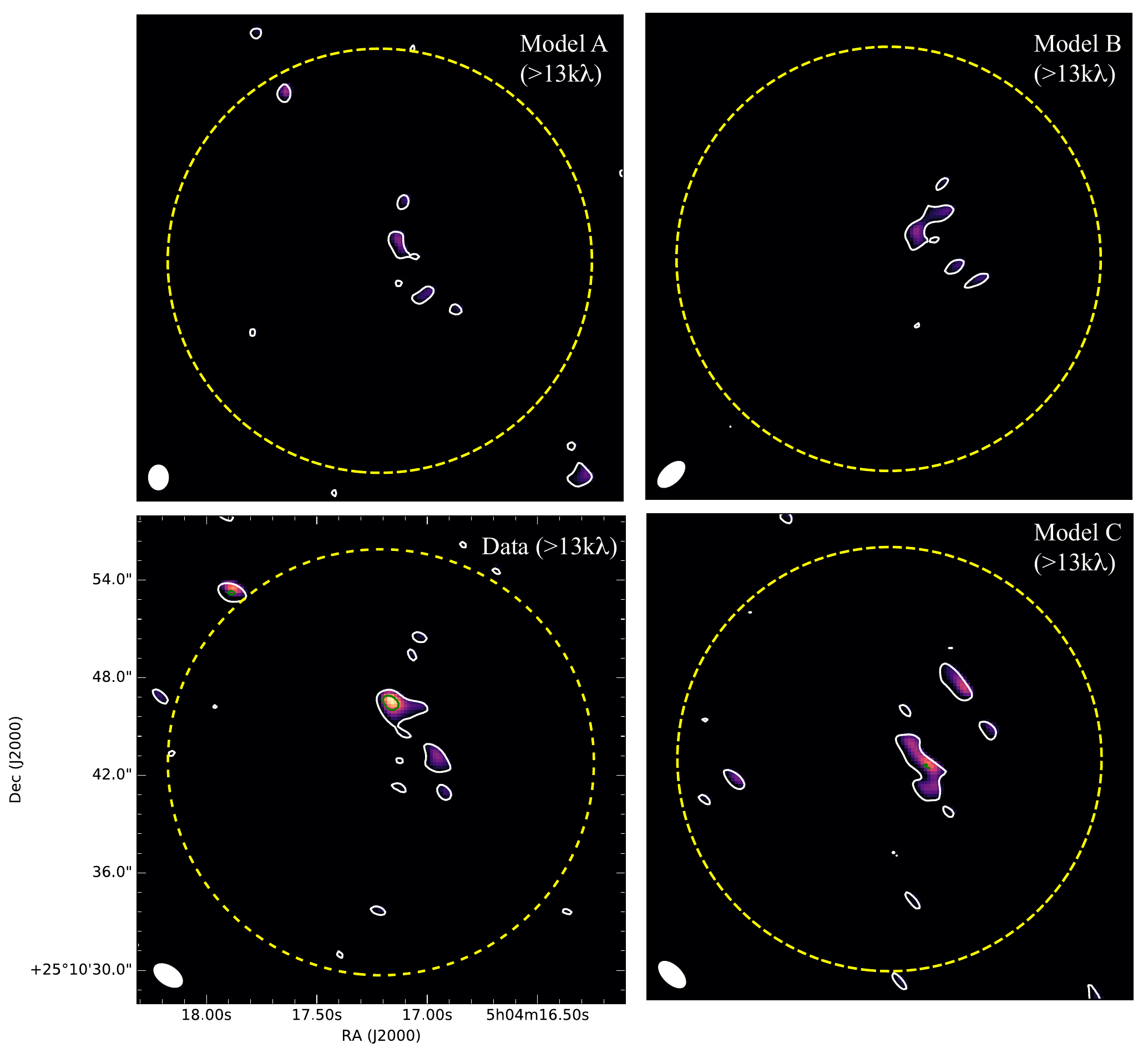} 
\caption{Continuum images using the longer baselines ($>$13k$\lambda$) for the observed data (bottom left) and for three different synthetic observations (different hour angle for observations) of the same smooth model (see text for more details). In all shown synthetic observations panels we can see spurious compact substructure at a similar level as seen in the ALMA data. This shows that the structure observed could also be reproduced with a smooth radial profile model. Beam sizes are shown in the bottom left corners for each panel. \label{fig:large_kilolambda}}
\end{figure}

We image both the real data and the synthetic ones with the same CLEAN parameters, but since we want to determine how reliable is the identification of small scale structure identified in the image, we only use baselines longer than 13\,k$\lambda$ and natural weighting. The resulting set of images have similar beams and a selection of those reproducing substructure similar to that observed is shown in Figure\,\ref{fig:large_kilolambda}.  They show that small substructures similar to those identified in the ALMA data of L1544 can be produced when a radial emission profile with a flat central section is used as input image, thus confirming that incomplete cancellation of Fourier components (sidelobes) could produce it.  

\subsection{Structure derived from single dish observations} \label{sec:chacon}
\begin{figure}[ht!]
\includegraphics[width=\textwidth]{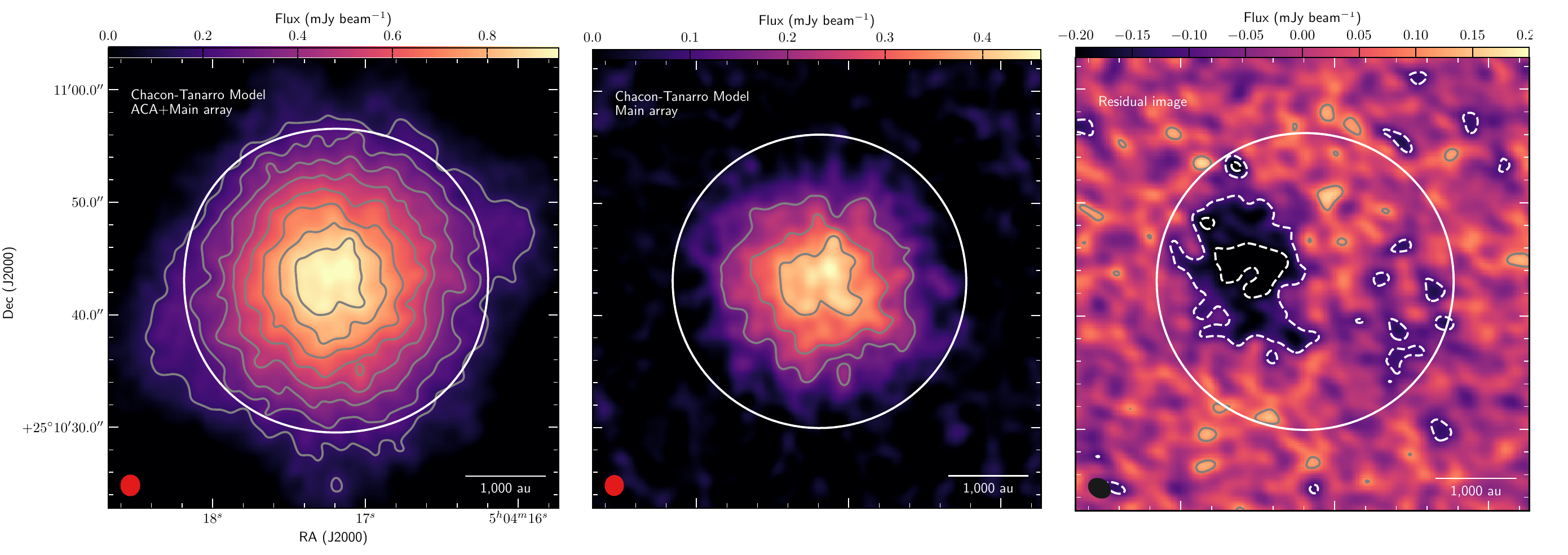} 
\caption{Results of simulating a model cloud with a density distribution obtained from the high-sensitivity single-dish observations of \citet{CPC19}. The color stretch and contour levels of the left and center panel are exactly the same (in Jy\,beam$^{-1}$) as those used for the data figures (starting from 5\,$\sigma$ and in steps of 3\,$\sigma$, with $\sigma$ = 36\,$\mu$Jy\,beam$^{-1}$). The left panel shows the ACA+Main Array observation simulations (to be compared with the right panel of Fig.\,\ref{fig:maps_array}); the central panel shows the Main array observation simulations (to be compared with the central panel of Fig.\,\ref{fig:maps_array}); the right panel is the residual image after subtracting the simulated image in the central panel to the equivalent data image in Fig.\,\ref{fig:maps_array}. The contours in this last panel start from 3\,$\sigma$ and they are in steps of 1\,$\sigma$. The dashed contours represent negative 3-$\sigma$ contours.}
\label{fig:Chacon}
\end{figure}

As already seen in the previous sub-section, the presence of an extended flat region in the central 1000\,au could produce artefacts when viewed with an interferometer.  However, in the previous subsection, the peak flux of the simulated combined image was underestimated because the model was constructed from the ALMA data, which did not account for the more extended structure due to filtering. Recently, single-dish dust continuum emission observations of L1544 at 1.1\,mm, with the AzTEC camera at the Large Millimeter Telescope (LMT), and at 3.3\,mm, with the MUSTANG-2 camera at the Green Bank Telescope (GBT), have been carried out toward L1544 \citep{CPC19}. The beam sizes are 12.6$^{\prime\prime}$ (corresponding to a linear size of 1701\,au) at 1.1\,mm and 9.7$^{\prime\prime}$ ($\equiv$1310\,au) at 3.3\,mm. These observations provided an updated density profile of L1544, which has been found to be similar to the one already proposed by \citet{CCW07}, taking into account dust opacity variations, which are found to be consistent with the presence of bare grains in the outer part of the core and the presence of dust grains with thick icy mantles toward the central 2000\,au \citep{CPC19}:

\begin{equation}
n(r) = \frac{n_0}{1+(r/R_{\rm flat})^{\alpha}}, \label{eq:tafalla}
\end{equation}

\noindent
where $n(r)$ is the number density of H$_2$ at radius $r$, $n_0$ is the central number density. The parameters found by \citet{CPC19} are: $n_0$ = 1.6$\times$10$^6$\,cm$^{-3}$, $R_{\rm flat}$ = 17.3$^{\prime\prime}$ ($\equiv$2336\,au), and $\alpha$ = 2.6 \citep[we note that the density profile in Eq.\,\ref{eq:tafalla} reproduces well the density profile of a Bonnor-Ebert sphere, when $\alpha$ = 2.5, as noticed by][]{TMC04}. Although these single-dish observations cannot resolve the L1544 kernel, we use this structure to test if it can reproduce our ALMA observations.   Figure\,\ref{fig:Chacon} shows the ALMA simulated observations of the pre-stellar core with the spherically-symmetric physical structure derived by \citet{CPC19}. From Figure\,\ref{fig:Chacon}, it is interesting to see that the core structure closely resembles that observed by ALMA (12m array + ACA), including the flux in the ACA + 12m and 12m-only images, as well as the apparent fragmentation in the central panel, although not at a significant level as in the central panel of Fig.\,\ref{fig:maps_array}. The large black area in the residual image in the right panel of Fig.\,\ref{fig:Chacon} is due to the presence of extra emission in the model cloud in the East side, as the model is spherical, while the data clearly show elongated structure (with a peak in the West side of the primary beam). 

\section{Hydro- and Magneto-hydro simulations of contracting pre-stellar cores} \label{sec:modeling}
Density perturbations within starless cores likely dissipate before becoming gravitationally unstable, as it has been shown by \citet{TGV18} analytically and also with three-dimensional (3D) hydrodynamic simulations. These authors suggest that solenoidal modes may eventually promote fragmentation via formation of vortical structure, but vortices can be disrupted by magnetic fields due to magnetic stresses. To investigate further this point, we carry out 3D non-ideal magneto-hydrodynamic (MHD) turbulent simulations using ZeusTW code \citep{Krasnopolsky+2010} to model the evolution of the pre-stellar core L1544. We adopt the same chemical network as in \citet{ZCL16} for obtaining the magnetic diffusivity of ambipolar diffusion (AD).

We first evolve under AD a uniform, spherical, magnetised core with total mass $M_{\rm c}=8.1~M_{\sun}$ \citep[similar to the total mass of the L1544 core within the C$^{18}$O(1--0) contours;][]{TMM98} and radius $R_c=5\times 10^{4}$~au, which corresponds to an initial number density for molecular hydrogen $n({\rm H_2}) \approx 2.3\times10^4$~cm$^{-3}$. The free-fall time of the core is $t_{\rm ff} \approx 7\times10^5$~yr. The initial core is threaded by a uniform magnetic field along the rotation axis with a constant strength $B_0 \approx 14.8~\mu$G \citep[consistent with][]{CrutcherTroland2000}, which gives a dimensionless mass-to-flux ratio $\lambda$ ($\equiv{M_{\rm c} \over {\pi R_{\rm c}^2 B_0}}2\pi\sqrt{G}$) of unity. The initial condition of the first stage is similar to that adopted by \citet{CiolekBasu2000}, therefore, AD is necessary to allow the magnetised core to collapse, instead of oscillating in the radial direction.

After the core evolves under AD for $\sim$0.7~Myr into a centrally concentrated structure with central number density of $5\times10^{4}$~cm$^{-3}$, we perturb the core with a $\mathcal{M}\approx5$ turbulence of Kolmogorov type (with spectral power law exponent -11/3). A small amount of solid body rotation with $\omega=4\times10^{-14}$~s$^{-1}$, as measured by \citet{Ohashi+1999},  is also added to the core. Due to the strong magnetic support in the critical core, it is difficult for the turbulence motions to compress the flow compared to the frequent formation of shocks in non-magnetised clouds. As a reference, we also perform a non-magnetised model (HD) for the same core, in which clumps are easily formed by turbulent compression. The core in the HD case also flattens along to the angular momentum direction, due to angular momentum conservation in the rotating cloud; however, the thickness of the flattened region is larger than the MHD model, which results in higher column density than the MHD model for a similar central number density.

\begin{figure}[ht!]
\plotone{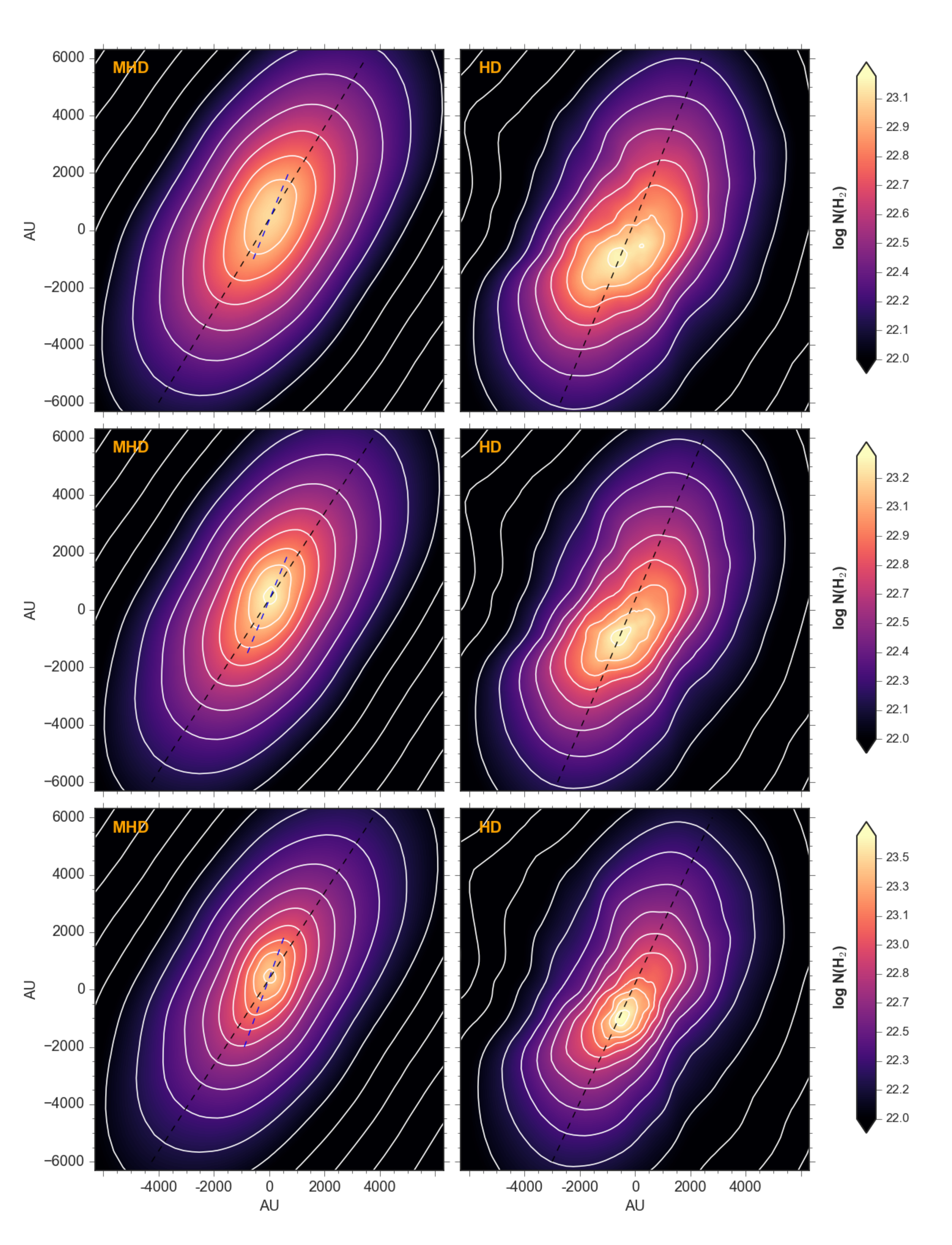}
\caption{Distribution of logarithmic column density (cm$^{-2}$) within the radius 6300~AU ($\sim$45\arcsec) of the modelled core, when the central density reaches 2$\times$10$^6$~cm$^{-3}$ (top panels), 5$\times$10$^6$~cm$^{-3}$ (middle panels) and 1$\times10^7$~cm$^{-3}$ (bottom panels). Left panels are non-ideal MHD models with AD and right panels are non-magnetised models. The black and blue dashed lines in the left panels denote the major axis of the large scale ($\sim$6000\,au) core and the small scale (1000\,au) kernel, respectively, indicating the misalignment between the major axis measured by \cite{CCW07} at core scales and the major axis of the kernel measured with ALMA. The observed misalignment is not reproduced with HD simulations.}
\label{Fig:Column6pan}
\end{figure}

In Fig.~\ref{Fig:Column6pan}, we show for both the non-ideal MHD and HD models three frames when the peak number density reaches 2$\times$10$^6$~cm$^{-3}$, 5$\times$10$^6$~cm$^{-3}$ and 1$\times10^7$\,cm$^{-3}$, respectively. At such stages, the root-mean-square velocities are subsonic for both models, with $\mathcal{M}$ about 0.8 for the HD model and 0.5 for the MHD model, respectively, mainly dominated by the bulk infall motions. The central $\sim$10000~AU of the core has the shape of an oblate ellipsoid. We need to tilt the core by $\theta$=45$^\circ$ to match the projected axial ratio (q$\sim$0.45--0.5) in the plane of sky \citep{CCW07} in the 5000~AU scale. This inclination angle is larger than the 16$^\circ$ derived in \citet{CiolekBasu2000}, because the core is not a perfect slab and the thinner edges of the oblate ellipsoid contribute less to the column density. No obvious substructures are present in the projected column density maps (obtained by direct line-of-sight integration of the number density) in Fig.\,\ref{Fig:Column6pan}. It is interesting to note that only simulated cores with peak number densities of 1$\times10^7$\,cm$^{-3}$ reach average number densities as high as those derived by \citet{CPC19} within 2336\,au (the flat radius found in their work; see Section\,\ref{sec:chacon}), thus indicating that single-dish observations are consistent with the presence of higher density regions, more compact than the telescope beam, within the flat radius. 

\begin{figure}
\plottwo{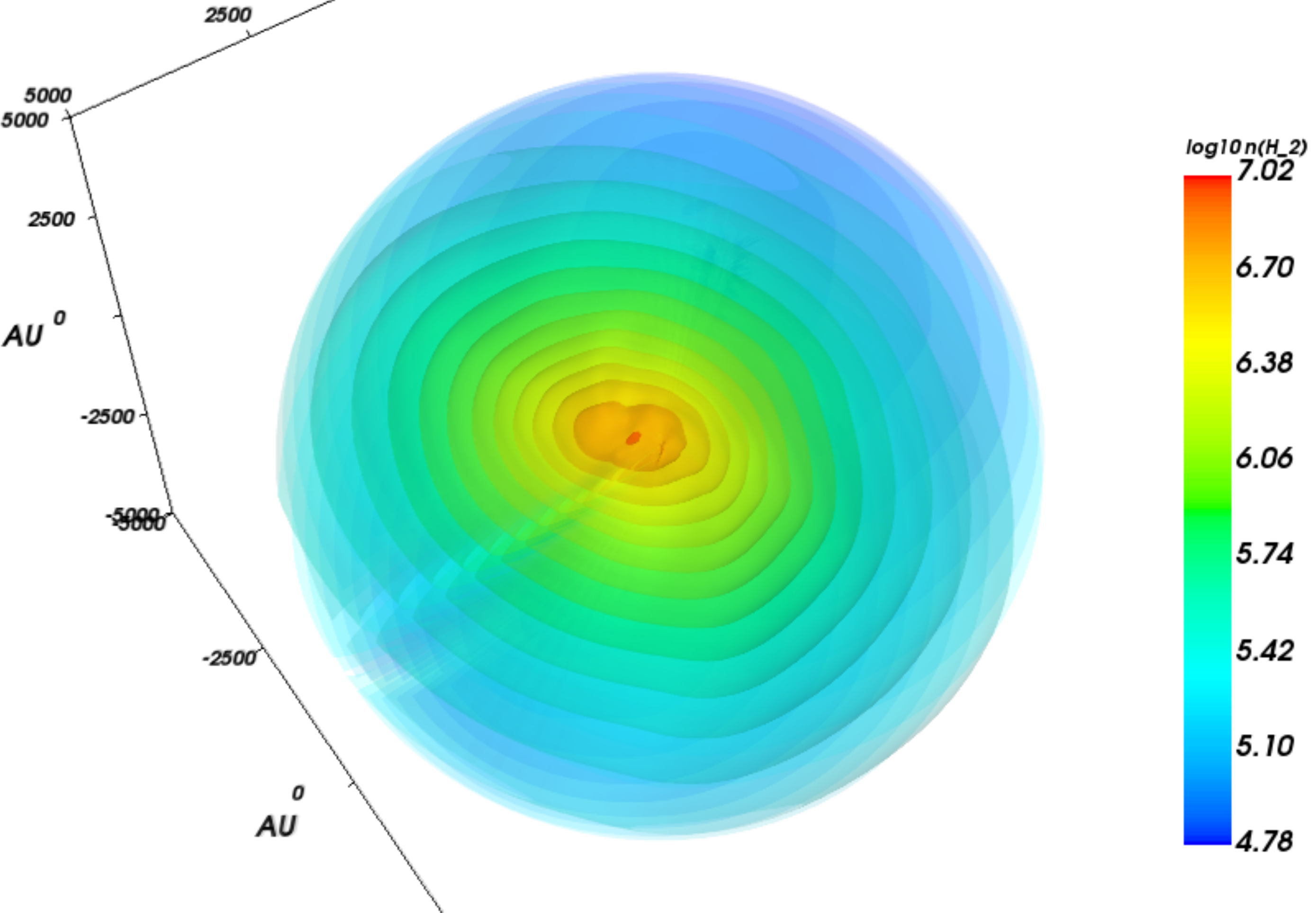}{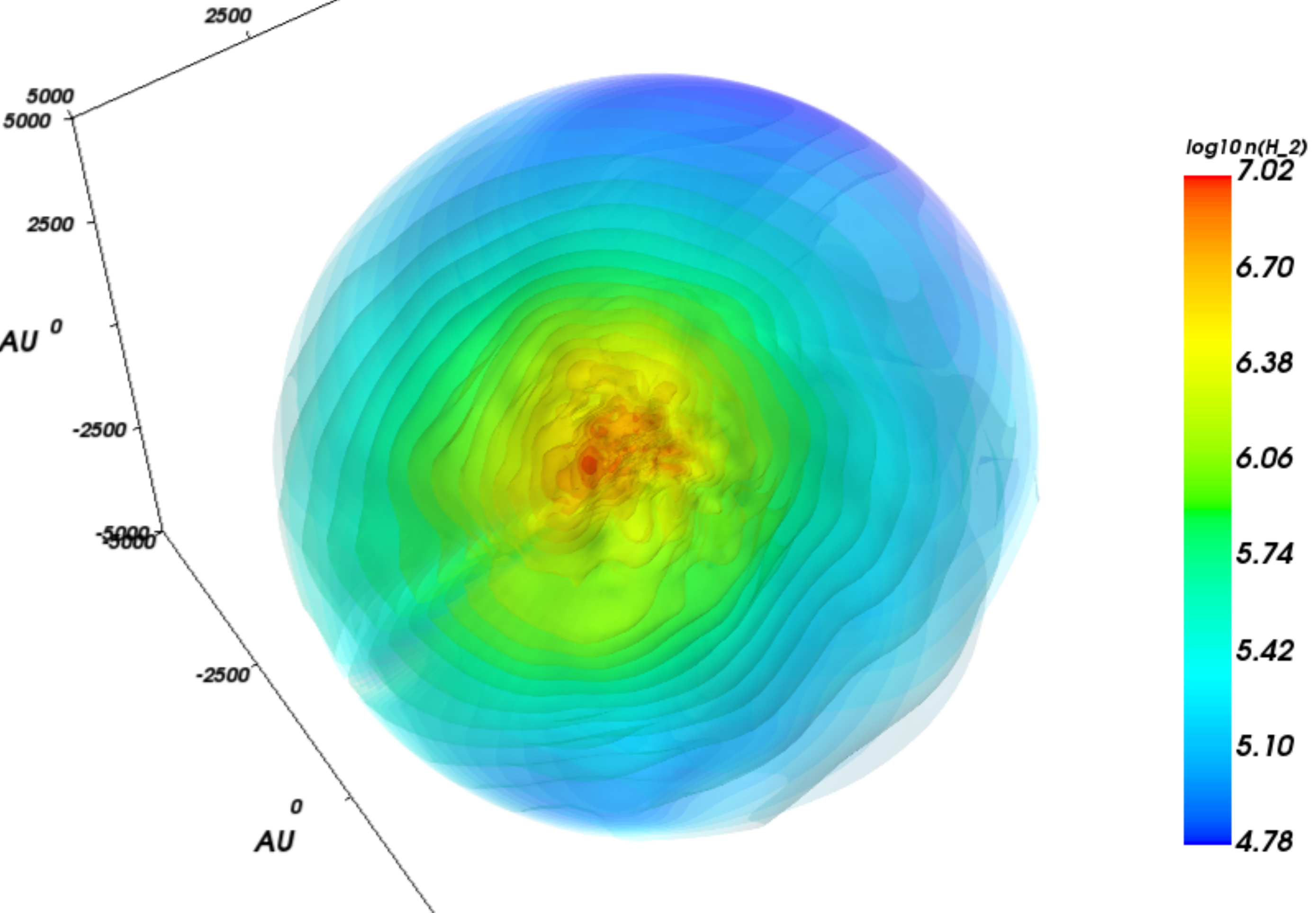}
\caption{3D number density maps of the MHD (left) and HD (right) simulated cores with central peak densities of 1$\times10^7$\,cm$^{-3}$. The colour scale refers to the logarithm of the H$_2$ number density, as labelled on the right of each panel. Note that the HD core has significant more substructure compared to the MHD core. However, the substructure is not distinguishable in the projected column density maps shown in Fig.\,\ref{Fig:Column6pan} as well as in the simulated ALMA observation images shown in Fig.\,\ref{Fig:obs_sim}.}
\label{fig:3D_number_density}
\end{figure}

In the 3D number density map,  the central region in the HD model is clearly more clumpy and warped than its MHD counterpart, in agreement with previous work \citep[e.g.][]{CLM12}; the peak of the gravitational potential in the HD model is also shifted away from the cloud centre. These additional asymmetric structures are developed in the HD model because of the flow's susceptibility to turbulent compression. In the projected maps of Fig.\,\ref{Fig:Column6pan}, it is difficult to visualise the structure of the core kernel, so we display in Figure\,\ref{fig:3D_number_density} the 3D number density maps of the MHD and HD cores which reach peak number density of 1$\times10^7$\,cm$^{-3}$. This clearly shows that the HD simulation produces substructure within the kernel, while the MHD simulation produces a smooth structure all the way down to the core centre. 

\begin{figure}
\plotone{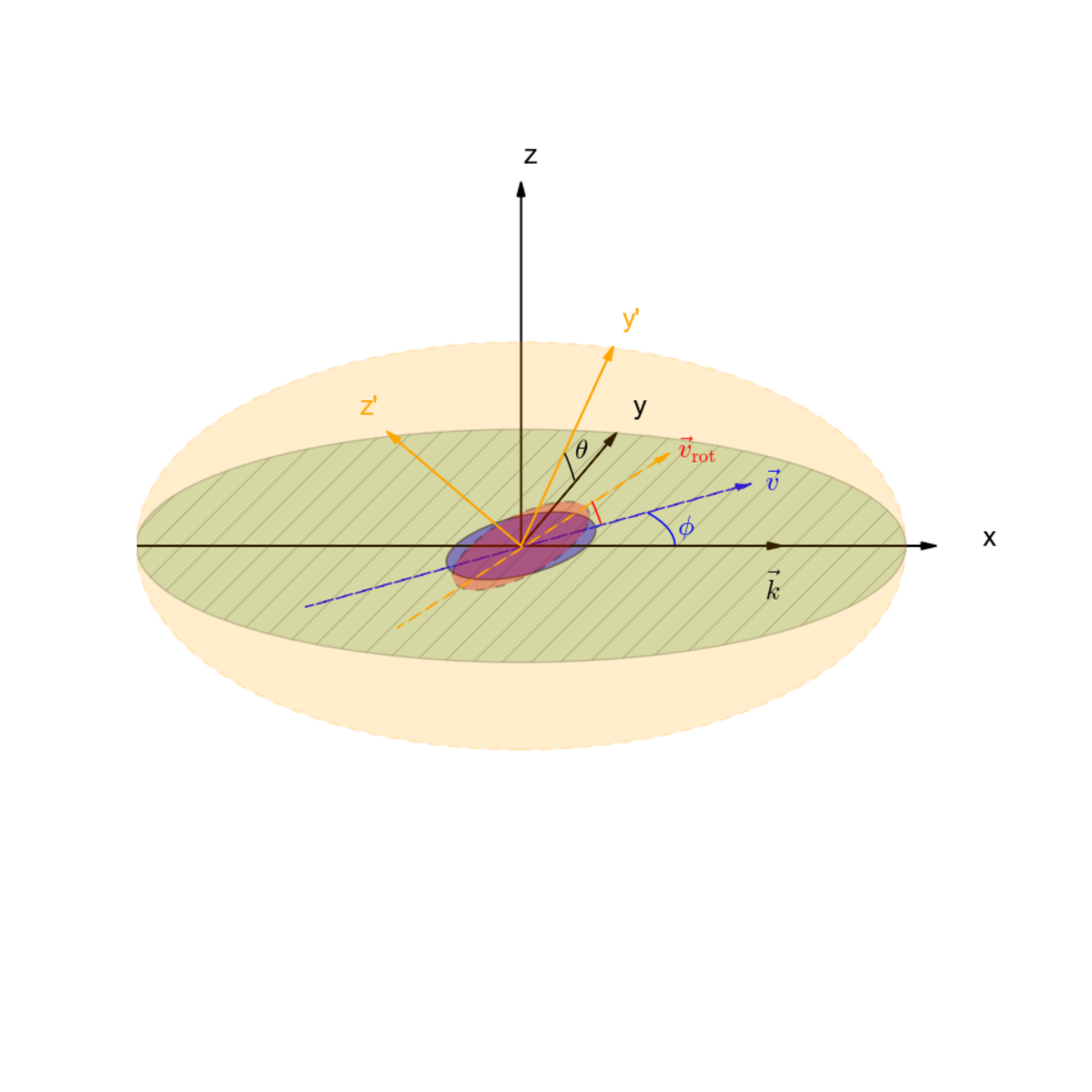}
\caption{Illustration of the rotation of the elongated kernel around axis $\vec{k}$. Green shaded region represents the original ellipsoid before tilting, and orange represents the ellipsoid after tilting by angle $\theta$. The original elongated kernel with major axis along $\vec{v}$ are hence rotated into the kernel with major axis $\vec{v}_{\rm rot}$. Such a misalignment will disappear if $\vec{v}$ is parallel to $\vec{k}$ or when the kernel has a similar axis ratio as the larger oblate core.}
\label{Fig:sketchRot}
\end{figure}

Another noticeable feature of the kernel in our ALMA map is the misalignment between its major axis and that of the NH$_3$(1,1) map observed by \citet{CCW07} at a slightly larger scale (90\arcsec; see left panels of Fig.\ref{Fig:Column6pan}). Our MHD models are able to reproduce such a misalignment, which is a natural outcome of rotating an elongated kernel (having a non-zero angle $\phi$ with the tilting axis) around the tilting axis of the oblate ellipsoid.  The original orientation angle $\phi$ of the kernel can be obtained via the Rodrigues rotation formula, explained in the following text and with the help of Figure\,\ref{Fig:sketchRot}. Assume that vector ${\bf v}$ = (cos$\phi$, sin$\phi$, 0) represents the original orientation of the kernel in the plane of the oblate ellipsoid, and vector ${\bf k}$ = (1, 0, 0) represents the tilting axis around which the ellipsoid rotates. After inclining the ellipsoid by an angle $\theta$, ${\bf v}$ also rotates to a new vector ${\bf v}_{\rm rot}$ = (cos$\phi$, sin$\phi$~cos$\theta$, sin$\phi$~sin$\theta$), whose projection on the plane of sky ${\bf v}_{\rm proj}$ = (cos$\phi$, sin$\phi$~sin$\theta$) can be related to the misalignment angle $\delta$ between the major axis of the kernel and that of the nucleus. Therefore, tan$\phi$ = tan$\delta$/sin$\theta$, which yields $\phi\approx45^{\circ}$ for the kernel of L1544. This implies that the kernel is by itself an elongated substructure that is well-distinguished from the larger oblate core. 

\subsection{Simulated ALMA observations}
\begin{figure}[ht!]
\plotone{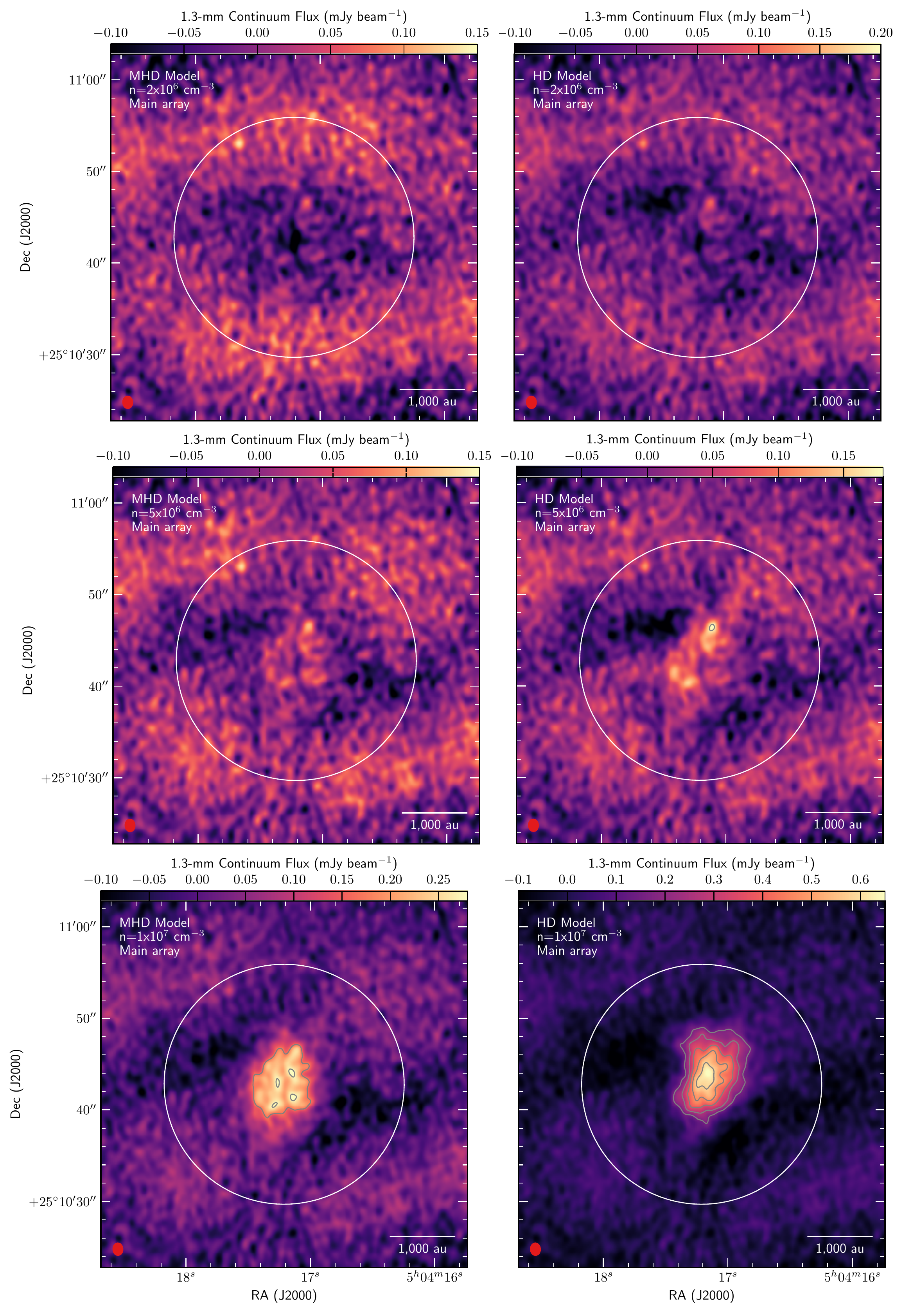}
\caption{Simulated ALMA Main Array observations of the cores in Fig.\,\ref{Fig:Column6pan}, with MHD simulations in the left and HD simulation in the right panels.  Significant emission is only found when the central densities are 1$\times10^7$\,cm$^{-3}$, due to the fact that at lower central densities the dense cores have smooth and large inner kernels, which are filtered out by the Main Array. In the panels with detections, contours start at 5\,$\sigma$ and are in steps of 3\,$\sigma$ (with $\sigma$ = 36\,$\mu$Jy\,beam$^{-1}$). The white circle is the primary beam and the red ellipse in the bottom left corner is the synthesized beam.}
\label{Fig:obs_sim}
\end{figure}

We first apply radiative transfer to the simulation models using RADMC-3D \citep{DJP12}. The temperature distribution of the core, ranging from $\simeq$5\,K to $\simeq$15\,K, is computed taking into account the external illumination from the interstellar radiation field through RADMC-3D; the temperature in the central 3000\,au is $\simeq$6 K, consistent with the measured values \citep{CCW07}. Simulated ALMA Main Array observations of the HD and MHD cores in Fig.\,\ref{Fig:Column6pan}, following the same procedure as for previous simulated observations presented in Sections\,\ref{sec:smooth-model}, can be found in Fig.\,\ref{Fig:obs_sim}, where the panels have been centred at the peak flux positions of the cores in Fig.\,\ref{Fig:Column6pan}. The order of the panels in Fig.\,\ref{Fig:obs_sim} is the same as in Fig.\,\ref{Fig:Column6pan}, with the MHD cores in the left panels and HD cores in the right panels; the cores have central number densities increasing from top to bottom: 2$\times10^6$\,cm$^{-3}$, 5$\times10^6$\,cm$^{-3}$ and 1$\times10^7$\,cm$^{-3}$.  It is interesting to note that the less centrally concentrated cores (those with peak number densities 2$\times10^6$\,cm$^{-3}$ and 5$\times10^6$\,cm$^{-3}$) are completely filtered out by the Main Array because of their smooth structure, while the cores with peak number densities of 1$\times10^7$\,cm$^{-3}$ can be well detected, both in the MHD and HD cases. This could explain why detecting starless dense cores with interferometers has so far been unsuccessful (as reviewed in Sect.\,\ref{sec:intro}): only (short-lived) dynamically evolved cores on the verge of star formation (in our case, those with peak number densities larger than 5$\times$10$^6$\,cm$^{-3}$), can be detected, as also discussed by \citet{DOP16}. In particular, we note that the 10$^7$\,cm$^{-3}$ MHD core has a peak flux about 1.5 times lower than the observed core (see central panel of Fig.\,\ref{fig:maps_array}), while the 10$^7$\,cm$^{-3}$ HD core has a peak flux about 1.5 higher than observed. This implies that with the present dust continuum emission observations it is not possible to distinguish between MHD and HD models and that high sensitivity and high spectral resolution molecular line observations are needed to quantify the role of magnetic fields in the dynamical evolution of pre-stellar cores. The fact that the observed peak flux is close to that of the simulated cores with central peak density 10$^7$\,cm$^{-3}$ implies structural similarities with L1544, with the MHD simulation also reproducing the misalignment between kernel and dense core major axes. It is finally interesting to note that, on the one hand, the simulated observations of the MHD core with peak density $10^{7}$\,cm$^{-3}$ presents fragmentation at the three-sigma level within the central 1000\,au, despite a smooth model core (see Fig.\,\ref{fig:3D_number_density}), showing that the detection of fragments can be due to interferometric artefacts; on the other hand, the simulated observations of the HD core with peak density $10^{7}$\,cm$^{-3}$ shows a smooth structure (see Fig.\,\ref{Fig:obs_sim}), despite a fragmented model core (see Fig.\ref{fig:3D_number_density}), due to the small size of the fragments, well embedded within the kernel. 

\section{Discussion and conclusion} \label{sec:discussion}

ALMA (Main Array + ACA) observations of the dust continuum emission at 1.3\,mm of the 8 M$_{\odot}$ pre-stellar core L1544 have revealed a kernel (defined by the 5$\sigma$ contour) with radius of 1446\,au, mass of 0.1\,M$_{\odot}$ and average H$_2$ number density of 1$\times10^6$\,cm$^{-3}$. Within the kernel, two fragments at the 3-$\sigma$ level in the Main Array image, and three fragments at the 2-$\sigma$ level in the Main Array+ACA image, are found. The fragments, if real, are unresolved structures, with radii $\leq$92\,au, mass $\leq$1\,M$_{\rm J}$, and peak volume densities $\simeq$5$\times10^6$\,cm$^{-3}$; they cannot be close to gravitational collapse. 

The comparison of our ALMA data with simulated observations of various smooth cloud models shows that: (1) the \citet{KRC14} Bonnor-Ebert sphere is more centrally concentrated, with central fluxes about two times larger, than observations; (2) in the case of the smooth cloud model with structure deduced from our ALMA observations, substructure similar to the observed fragments appears in the simulated observations, confirming that incomplete cancellation of Fourier components can produce the observed fragments; (3) when the latest (spherically symmetric) physical structure of the L1544, deduced from single dish observations, is used as input in the simulated observations, a good match with the peak fluxes in the ACA+Main Array and Main Array-only images is obtained, although substructure at the observed $\sigma$ levels is not reproduced. 

Non-ideal MHD and HD simulations of a dense core similar to L1544 have been run until the central density of the core reaches peak number densities 2$\times10^6$\,cm$^{-3}$, 5$\times10^6$\,cm$^{-3}$ and 1$\times10^7$\,cm$^{-3}$.  The MHD simulated cores have a smooth structure and present a misalignment between the major axis of the dense core and that of the kernel, similar to that found with observations. The HD simulated cores present substructure and warped kernels. When ALMA Main Array simulated observations are carried out, only model cores with peak number densities of 1$\times10^7$\,cm$^{-3}$ can be detected. These cores have average number densities, within a flattening radius of $\simeq$2300\au, close to those deduced by recent single dish observations \citep[$\simeq$1.6$\times10^6$\,cm$^{-3}$;][]{CPC19}. Thus, only short-lived dynamically evolved and centrally concentrated pre-stellar cores can be studied at high angular resolution; this explains why it has been so far difficult to detect starless cores with interferometers, including ALMA.  The simulated observation of the high density MHD core shows substructure within the kernel at levels similar to the observed ones, despite the smooth model core; this again shows that interferometric artefacts can mimic fragmentation. The simulated observations of the high density HD core shows a smooth structure despite the model core being highly fragmented toward the central region; this is due to the small size of the fragments compared to the synthesized beam, suggesting that even higher angular resolution observations should be carried out to test MHD and HD predictions.  In the future, we will compare the simulated and measured velocity fields to investigate more quantitatively the non-ideal MHD and HD predictions (Pineda et al., in prep.). 

With the current data, we cannot claim evidence of fragmentation. Nevertheless, fragmentation can take place.  The Taurus Molecular Cloud is known to host a large number of binary pairs: between 2/3 and 3/4 of the whole population of young stellar objects are multiple systems of two or more stars, while the rest appears to have formed as single stars \citep{KIM11}. The separation of the observed stellar multiples shows values around a few hundred au, thus indicating that fragmentation may happen within the flat central region of L1544. Basically, star formation does happen and it tends to produce multiple systems. Core fragmentation is one of the main expected modes of binary/multiple system formation \citep{RCB14}, but the present data, toward one of the most massive and dynamically evolved starless dense cores, do not support this scenario, suggesting that either L1544 will form a single star or that fragmentation may happen at a later stage of evolution, maybe soon after the formation of the first protostar \citep[see e.g.][]{POP15}. Alternatively, fragmentation could indeed happen at pre-stellar stages, but higher sensitivity observations are needed to test this scenario. Our data are also pointing out that caution should be taken when claiming fragmentation within pre-stellar cores with ALMA.

\acknowledgments

We acknowledge the very helpful comments of our referee, Neal Evans, and Anika Schmiedeke for her help on the radiative transfer. P.C., J.E.P., B.Z. and A.C.-T. acknowledge the financial support of the European Research Council (ERC; Advanced Grant PALs 320620).  MP acknowledges funding from the European Unions Horizon 2020 research and innovation programme under the Marie Sk\l{}odowska-Curie grant agreement No 664931.


\vspace{5mm}
\facilities{ALMA}


\software{ZeusTW; \citep{Krasnopolsky+2010,ZCL16},  
          RADMC3D \citep{DJP12}, 
        CASA \citep{MWS07}
                }

\end{document}